\documentclass[10pt, conference, compsocconf]{IEEEtran}
\IEEEoverridecommandlockouts
\usepackage{cite}
\usepackage{amsmath,amssymb,amsfonts}
\usepackage{algorithmic}
\usepackage{graphicx}
\usepackage{textcomp}
\usepackage{comment}
\usepackage{enumitem}
\usepackage{subfigure}
\usepackage{textcomp}
\usepackage[table,xcdraw]{xcolor}
\usepackage{hhline}

\def\BibTeX{{\rm B\kern-.05em{\sc i\kern-.025em b}\kern-.08em
    T\kern-.1667em\lower.7ex\hbox{E}\kern-.125emX}}
\begin{document}

\title{MAC Protocol Design Optimization Using Deep Learning\vspace*{-0.1in}}

 
\author{\IEEEauthorblockN{Hannaneh Barahouei Pasandi, Tamer Nadeem}
\IEEEauthorblockA{{Dept. of Computer Science, Virginia Commonwealth University, Richmond, VA 23284, USA} \\
\{barahoueipash, tnadeem\}@vcu.edu
}}
\maketitle

\begin{abstract}
Deep learning (DL)-based solutions have recently been developed for communication protocol design. Such learning-based solutions can avoid manual efforts to tune individual protocol parameters. While these solutions look promising, they are hard to interpret due to the black-box nature of the ML techniques. To this end, we propose a novel DRL-based framework to systematically design and evaluate networking protocols. While other proposed ML-based methods mainly focus on tuning individual protocol parameters (e.g., adjusting contention window), our main contribution is to decouple a protocol into a set of parametric modules, each representing a main protocol functionality and is used as DRL input to better understand the generated protocols design optimization and analyze them in a systematic fashion. As a case study, we introduce and evaluate DeepMAC a framework in which a MAC protocol is decoupled into a set of blocks across popular flavors of 802.11 WLANs (e.g., 802.11a/b/g/n/ac). We are interested to see what blocks are selected by DeepMAC across different networking scenarios and whether DeepMAC is able to adapt to network dynamics. 

\end{abstract}

\section{Introduction and Motivation}

The proliferation of the current Internet and mobile communications networked devices, systems, and applications has contributed to increasingly large-scale, heterogeneous, dynamic, and systematically complex networks. The increasing availability and performance demands of these applications suggest that "general-purpose" protocol stacks are not always adequate and need to be replaced by application tailored protocols. The current approach of protocol design is to preprogram empirical control rules with configurable thresholds that can be adjusted by domain experts in an iterative trial-and-error manner on the field for each scenario \footnote{In this paper the term scenario refers to different scenarios included in user objective, device constraint, network condition, and application requirements.}. However, heterogeneity and dynamically changing characteristics of networks (e.g., IoT) ask for intelligent, and auto-configurable techniques. In this era, network protocol design requires a  new approach in which control rule optimizations are not only based on closed-form analysis of isolated protocols, but are based on high-level policy objectives and a comprehensive view of the underlying components.

\begin{figure}[!t]
	\centering
	\includegraphics[width=0.48\textwidth,height=1.9 in]{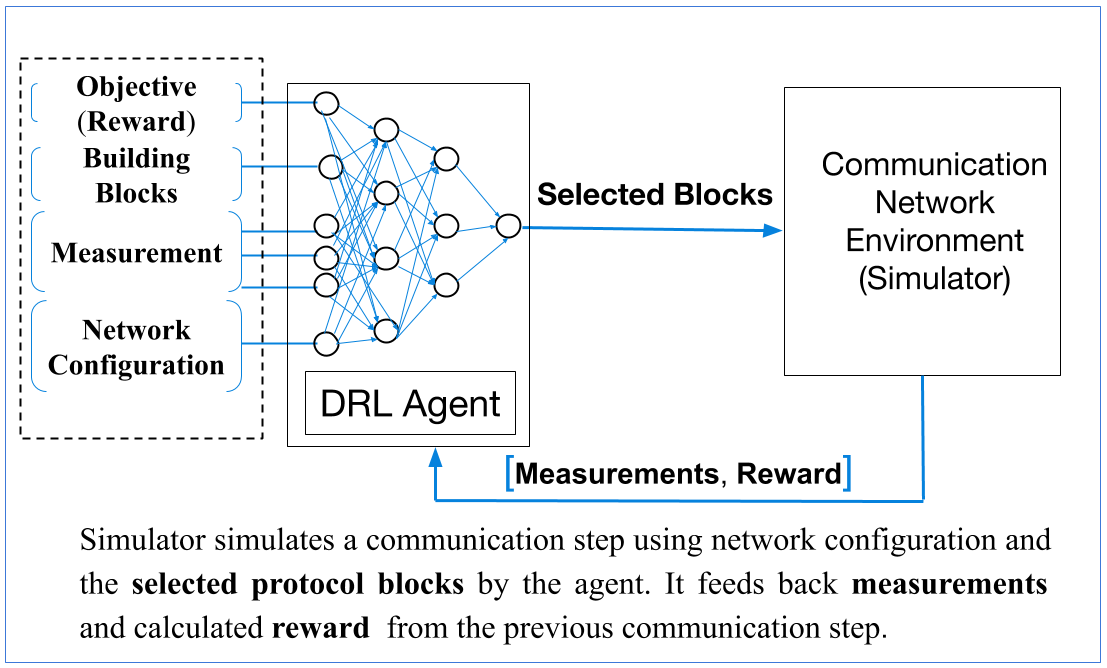}
	\caption{A Bird's-Eye View of the Proposed Framework } 
	\label{fig:eagle}
\end{figure}

Cutting-edge advances in deep reinforcement learning (DRL) have recently stirred up the network research community, now rushing to apply DRL to various protocol optimization tasks such as routing~\cite{geyer:2018}, congestion control \cite{deepcc} and MAC protocol~\cite{dmdl}, just to name a few. Applying DRL techniques can reduce manual human-based efforts to tune protocol parameters. 

In this paper, we propose a DRL-based framework to optimize the design of networking protocols. As a case study, we narrow down our focus to propose a DRL-based framework for designing MAC protocols hereafter {\em DeepMAC}. In DeepMAC framework,  MAC protocols are decoupled into a set of parametric modules, each representing a main functionality across popular flavors of 802.11 WLANs (IEEE 802.11 a/b/g/n/ac amendments). The decoupled parametric modules, referred as {\em building blocks}, are used as DRL inputs (see Figure~\ref{fig:eagle}). Protocol designers provide only high-level specifications for a scenario, including the communication objective, the protocol building blocks, measurements, and network configuration. The DRL agent is then able to learn what protocol building blocks (functionality) are important to be included or to be neglected in the protocol design. As we showcase in Section~\S\ref{important-blocks}, the DRL agent learns that when the load of the network is very low, it could eliminate control and sensing mechanisms (ACK and Carrier Sensing blocks, respectively) to increase the throughput of the channel by reducing the bandwidth overhead and waiting time introduced in these mechanisms. Therefore, this framework could provide a tool for protocol designers to re-think the blocks used in a designed protocol. In addition, our framework can be utilized as a multi-variant optimization tool that helps in alleviating the current protocol design process. When designing a protocol, domain experts should keep different application requirements, user objectives, device constraint and network conditions in mind. Considering these parameters all together is a daunting task. Using this framework, domain experts could identify/capture the significance of components of protocol performance under varying scenarios for different objectives. It could also help them to get insights about the relation between different protocol components for different objectives, although such components may not have a direct dependency on each other if considered alone. More importantly, our framework goes beyond only {\em tuning} individual protocol parameters by proposing to decouple a protocol into its main functionalities to optimize the protocol design and to better understand the generated protocols and analyze them in a systematic fashion. Finally, merging different MAC protocols used for different technologies and devices is a common practice in the research community~\cite{sanitize:2017, queue:2012, piva:2019}. Our framework can also serve as a tool in which blocks (functionalities) extracted from different protocols (e.g., TDMA-like protocols used in Bluetooth or CSMA-like protocols used in IEEE 802.11 WLANs) can be used as the input to generate a new protocol. All-inclusive, the proposed framework provides the opportunity for protocol designers to explore a different combination of blocks across multiple protocols under different scenarios.  

A MAC protocol can be regarded as mapping a perceived history of feedback from the network condition, to the next choice of a set of building blocks for designing a protocol. We hypothesize that such history contains information (useful features) about network conditions that can be exploited for better protocol design by learning the mapping from experience via a DRL approach. We use our proposed building block-based method to demonstrate what blocks are selected by DeepMAC under different scenarios and why. We show that by using our approach, we develop a deeper understanding of the interplay between the parametric protocol blocks and the underlying environment.   

However, the main purpose of this paper is not to provide a comprehensive solution for the protocol design optimization and analysis problem using DRL techniques. The objective is to better understand the design optimization task systematically, beyond just looking at the bottom-line performance. We believe that this approach is not limited only to MAC layer protocols, but to the communication protocol stack as a whole.

\section{Related Work}
\label{RelatedWork}
In recent years, machine learning has shown excellent improvement across different fields~\cite{wic:2019, wic:2017, wick:2018, panahi:2019, dsa3, eini:2019}. We narrow down our focus to recent network research community's effort in exploiting DRL techniques to optimize PHY and  MAC protocols in mobile and wireless networks. However, deep learning for the wireless MAC sub-layer has been relatively much less explored. In the following, we summarize a few  recent works. Naparstek et al.~\cite{dsa} consider the problem of DSA for network utility maximization in multi-channel wireless networks. In their mechanism, the objective is a multi-user strategy for accessing the spectrum that maximizes a certain network utility in a distributed manner without online coordination between users. The action of each user at the beginning of each time slot is to {\em select} a channel and transmits a packet with a certain transmission probability. In~\cite{dsa2} the DSA problem is formulated as a partially observable Markov decision process (POMDP) with unknown system dynamics. In this framework, a user at each time slot {\em select} a channel to transmit data and receives a reward based on the success or failure of the transmission. Yu et al.~\cite{dlma} investigate a  DRL-based MAC protocol for heterogeneous wireless networks. In their model, DRL agent decides whether to transmit or to wait at each time slot with the objective of maximizing the sum throughput and $\alpha$-fairness among all networks. In our previous work,~\cite{pasandi:2019} we proposed an RL based framework to optimize the MAC protocol using a simple set of functionalities. However, we discovered that the RL-based approach may face instability since the agent has to find a balance between exploration and exploitation.  In~\cite{pasandi-b:2019, pasandi2019poster} we give a complete overview of the whole protocol design framework using machine learning techniques. In these works, we describe the key design considerations for the learning agent (e.g., centralized, distributed or hybrid agents) and describe how these agents should communicate with one another. We then expanded our framework~\cite{mac-unboxing:2020} to use deep architecture along with new building blocks.

Although the aforementioned mechanisms differ in the details, their common objective is to optimize a protocol is by \textit{tuning and/or controlling} the protocol parameters. Our proposed approach is different from these works in two significant respects. First, we argue that designing methods to boost protocol performance is not only about parameter tuning, but also to decide what functionality to include or exclude from the design. The novelty of our approach resides in the way our framework constructs a protocol from a set of building blocks. By decomposing the protocol into the set of mechanistic building blocks, we aim to better understand the design, the interdependencies among different protocol building blocks, and to ease the analysis of the protocols. In addition, our framework supports multi-variant communication objectives that can be \textit{explicitly} defined by domain experts. 

\section{DeepMAC Framework}
\label{framework}
\begin{figure}[t!]
	\centering
	\includegraphics[width=0.48\textwidth,height=2 in]{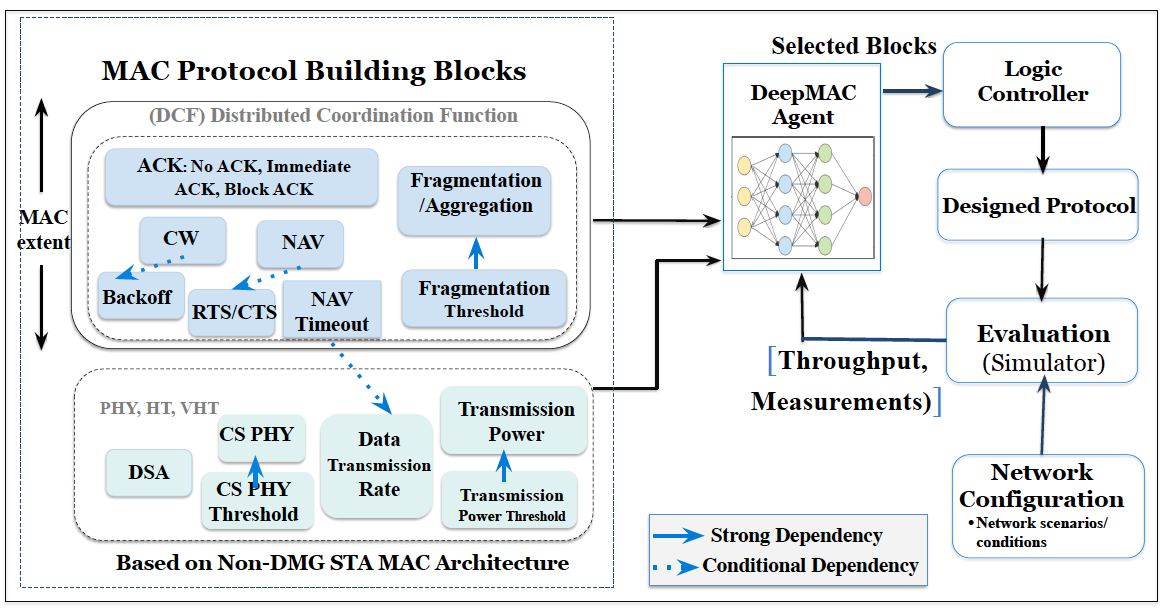}
		\vspace*{-0.1 in}
	\caption{DeepMAC framework}
	\label{fig:framework}
	\vspace*{-0.2 in}
\end{figure}

MAC protocols need to be designed with a rich set of requirements in order to satisfy the needs of the overlaying applications and scenarios. Due to the limited channel resources and a large number of devices accessing the channel, it is desirable that the MAC protocol minimizes the time wasted due to collisions or exchange of control messages. In addition, it is required that the effective throughput remain high irrespective of the traffic levels. Overwhelmingly, the network conditions may be dynamic (e.g., nodes entering and leaving). Thus, it is imperative that the MAC protocol can be easily scalable and adjusted delicately to the changing environment with little or no control information exchange. Figure~\ref{fig:framework} shows DeepMAC framework and its key modules that aim to optimize the design of wireless MAC protocols. We describe these modules in the following subsections. 

\vspace{-2mm}
\subsection{MAC Protocol Building Blocks}

A network protocol is typically structured into several layers, where each layer is broken into a set of protocol blocks with its own specific functionality. The modular design of protocols is a promising approach as acknowledged by~\cite{doerr:2005, bai:2004}. The building blocks are a set of separated parametric modular components, each of which is in charge of one (or several) specific well-defined functionality. The combination of different building blocks and the interactions between them determine the overall behavior of a network protocol for a given environment.

In our framework, we have extracted a set of MAC protocol blocks from Wireless LAN Medium Access Control (MAC) and Physical Layer (PHY) Specifications~\cite{ieee:2016} which includes MAC functionalities across all 802.11a/b/g/n/ac amendments. As shown in Figure~\ref{fig:framework}, these blocks, and instances of their dependencies are captured based on non-Directional Multi-Gigabit (non-DMG) MAC architecture. Having established a number of potential building blocks, DRL agent takes these blocks and history of the average channel throughput as its main inputs as described later. 

\subsection{DeepMAC As A Reinforcement Learning Problem}
DeepMAC uses Reinforcement Learning (RL) along with a deep architecture to learn the best set of protocol blocks for different scenarios. In DeepMAC, we consider a {\em centralized} learning agent for the design of 802.11 MAC protocols. This centralized agent, in practice, can be placed on a single supernode (e.g., the Access Point) that periodically updates its model. Meaning it decides the selected set of MAC layer blocks and parameters to be used by all the other nodes in the network.

\textbf{Reward function} The designers need to specify the communication objective according to corresponding use case,  device category under different scenarios. In the DRL framework, this defines the reward function. For example, for a battery-constrained IoT device which that cares for maximizing the throughput while minimizing spent energy, a designer may specify to optimize the following objective function: 

 $w_0$ (number of successful transmitted bits) - $w_1$ (energy spent per bit)

Where $w_0$ and $w_1$  control relative tradeoff between these two conflicting components.
In DeepMAC, the reward function is the \textit{average throughput} of the link. Although such reward objectives can change based on the provided scenario by the protocol designer.

\textbf{State, Action} The state of the agent is a vector of numerical representation of the set of the building blocks, and a history with a fixed link of the average link throughput values which are used as part of the input for DeepMAC agent. In this set, a value except 0 indicates that the corresponding block is included in the protocol design (each of the elements in the input vector can have different values which indicate what parameter or algorithm/method/mechanism should be used in the design), while 0 means the component is completely excluded from the design. The \textit{action} in this framework is the act of choosing the next state among all the available states from the current state such that the \textit{reward} is maximized.  Given the input, the output of the simulator is the average throughput of the channel which is considered as the reward of the DRL agent for the selected building blocks at the current step. We consider a global optimization of the reward function which relies on the assumption that all nodes employ the same prescribed protocol using the selected blocks by the agent. The agent takes both the protocol design blocks and history of the reward as inputs of the agent and outputs the best combination of building blocks for the current scenario that maximizes the reward. 

\textbf{Learning environment} The designers specify the network environment for which the DeepMAC must be optimized. In the DRL framework, this feeds into the \textit{learning environment}. For example, a designer might choose the reference model in~\cite{mac_model:2004} as a potential environment. Note that using such reference models is standard practice in the industry for designing and evaluating any new technologies as it offers a reproducible way for testing the technology in a wide variety of realistic and challenging deployment scenarios.

\textbf{DRL agent architecture} The neural network we adopt is equipped with three hidden layers and an output layer. We find through our experiments that this simple architecture can yield satisfactory performance, and increasing the complexity of the neural network does not contribute to performance improvements while inducing more training overload. The data is flattened before going through the hidden layers which utilize Relu as the activation function. The output layer consists of multiple neurons, each producing the Q-value of the corresponding action.

\subsection{Logic Controller}
In network protocols, some functional blocks are dependent on each other. The logic controller in our framework is designed to check a) the block execution sequences b) their interdependencies, and  c) interaction rules between blocks to ensure logically correct protocol design. Each protocol can be modeled as a directed graph where the vertices are the blocks and the edges are the interdependencies and interactions between these blocks. We extracted the interdependencies from PHY and MAC specification~\cite{ieee:2016}, and incorporated them in the logic controller using if-then-else rules.  All dependencies are uni-directional meaning if \textit{Block A} depends on \textit{Block B} it only shows restrictions of \textit{A} to \textit{B} but not \textit{B} to \textit{A}. We define two types of dependencies between blocks: \textit{Strong} and \textit{conditional} dependency. In our framework, a strong dependency is between those blocks that are tightly wired together and must be selected together in order to deliver their functionality properly. As an example, the Backoff mechanism is strongly dependent on the ACK block (Backoff$\,\to\,$ACK). Although the other direction does not hold, meaning we can use ACK without having a backoff mechanism.

In this framework, a conditional dependency is established based on the extracted rules from~\cite{ieee:2016}. We provide the following examples to describe some of such dependencies in the following. As shown in Figure~\ref{fig:framework}, NAV (virtual CS mechanism) is conditionally dependant on RTS/CTS block. There are two methods to set the NAV parameter: a) by reservation information distributed through the RTS/CTS Duration field and b) by information provided in the Duration/ID field in individually addressed frames. Therefore, in our framework, NAV is set based on the RTC/CTS block if this block is available (selected by the agent). Otherwise, it is set based on the latter approach. 

\subsection{Network Configuration and Designed Protocol Evaluation}
Integrated with protocol design element would be the network scenarios and conditions, such as communication medium types and node mobility. Different scenarios have different assumptions and requirements that need to be captured when designing a protocol. To evaluate DeepMAC framework, we developed an event-driven simulator using C++, while having the ns-3 design in mind. Our simulator mimics the MAC protocol of ns-3, but it is flexible to support the decomposed building blocks\footnote{ High-end simulation tools (such as Opnet, NS-3, etc.) have the ability of reproducing with an accuracy of implementation. However, such tools do not support our building block decomposition concept properly.}, and consequently the design of MAC protocols. Each building block is considered as a module and the agent decides about the inclusion and exclusion of the block as a part of protocol design. As an example, when the ACK mechanism is turned on, the agent may select either immediate or delayed ACK mechanism depending on the underlying network scenario. As input, the simulator takes the values of building blocks from the DRL agent that passed the logic controller check of finding any type of conflict or interdependency between them. It also receives the network configuration parameters including the number of nodes, level of noise, etc.

\section{DeepMAC Evaluation}
\label{evaluation}
\textbf{Performance Metrics} This section presents the numerical results and evaluation of DeepMAC regarding a) \textit{average throughput} enhancement and b) \textit{block selection} by the agent under different scenarios, respectively. Before we delve into the experimental evaluation of our analysis, we clarify that we run the pre-trained DRL agent for every scenario. After training our DRL agent on a MacBook Pro with 2.9 GHz Intel Core i5 with 16 GB of memory, the agent took on average 1 ms to execute. We assume that the supernode (centralized agent) uses hardware accelerators which can reduce the execution time by an order of magnitude and comfortably meet the real-time requirement. We have not considered the convergence time of the DRL agent as a performance metric to evaluate.

\begin{figure*}[!t]
\centering

\subfigure[]{\label{fig:lowload}\includegraphics[width=67mm]{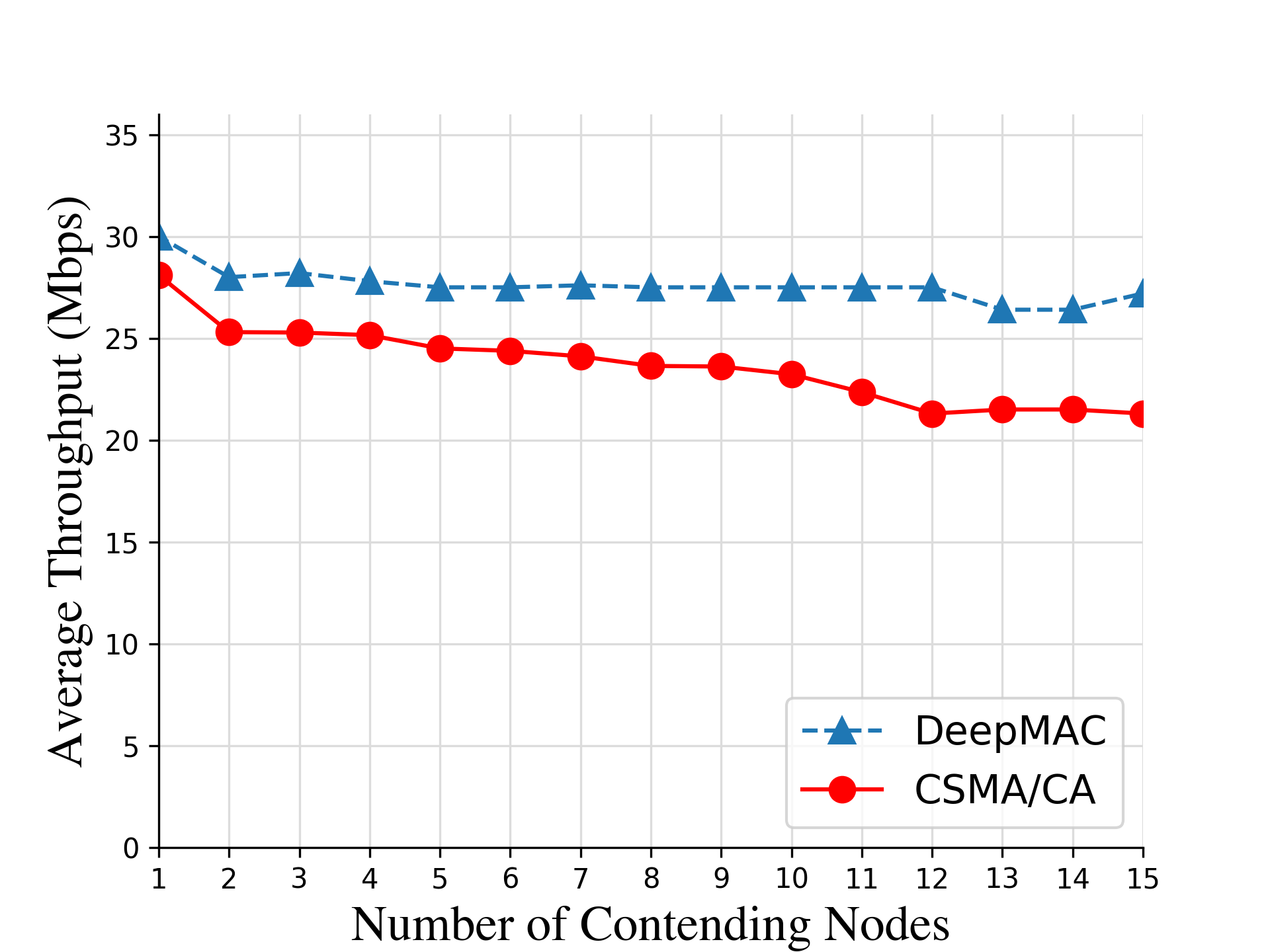}}
\subfigure[]{\label{fig:highload}\includegraphics[width=67mm]{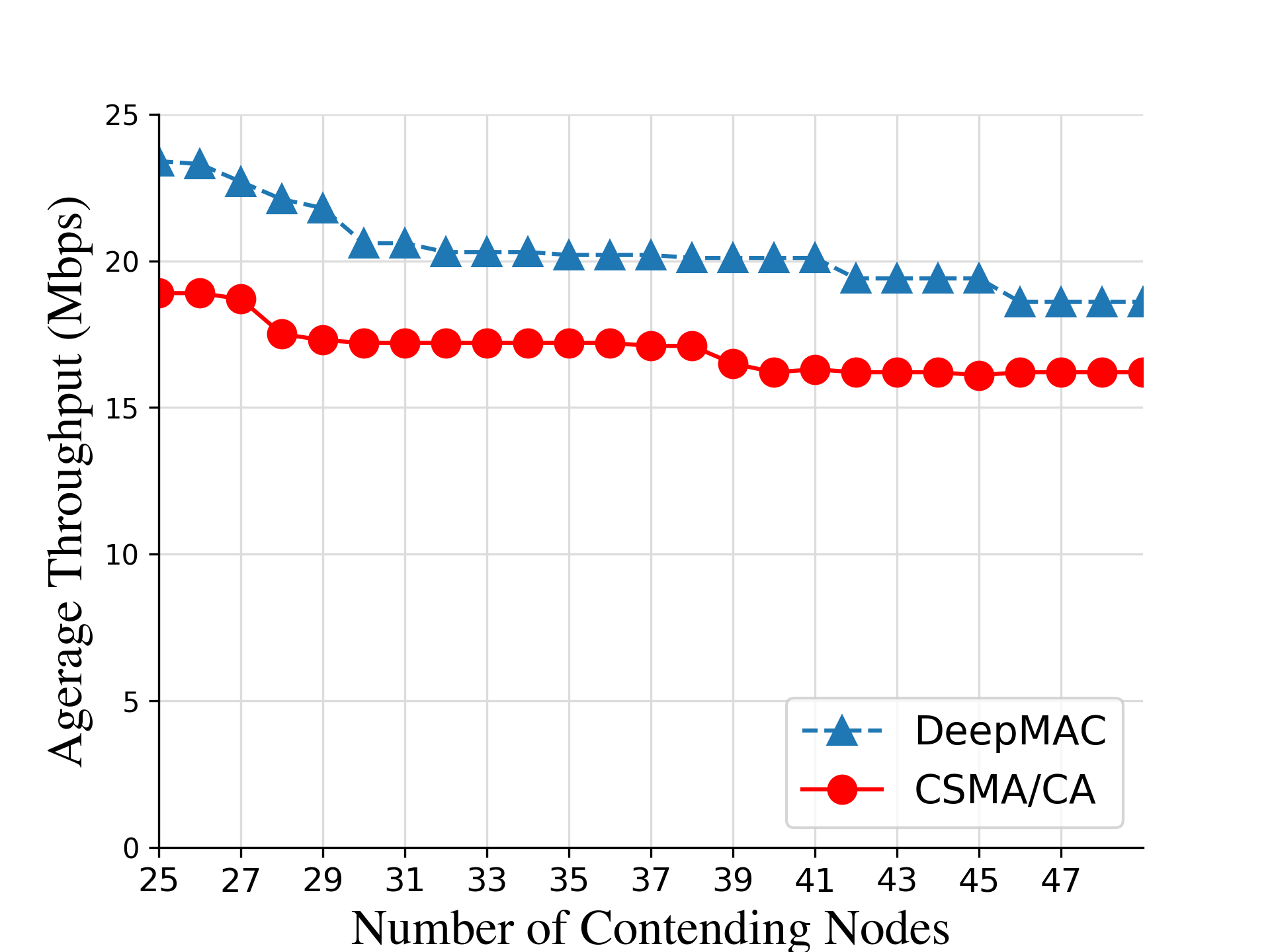}}

\caption{Throughput comparison of DeepMAC against CSMA/CA under \subref{fig:lowload}  Low load traffic. ~\subref{fig:highload} High load traffic.}
\label{th-compare}

\end{figure*}

\subsection{Simulation Configuration}

We consider an Ad-hoc network where individual nodes communicate with each other directly. To carry out our simulations, we use our event-driven simulator.  Table~\ref{table_sim} summarizes the simulation configuration parameters used in our experiments. The nodes are static and are randomly scattered in a 200x200m area. Without loss of generality, we assume that each node has always a packet to transmit, and the packet generation rate follows a Poisson process. In our experiments, we consider eight different networking scenarios described in Table~\ref{table_scenario}. The low load corresponds to an under-saturated network with 5 nodes, and average packet generation rate of 8 packets per second, while scenarios with high load represent close to saturated, and saturated networks with 20 to 50 nodes, and average packet generation rate of 470 packets per second. With regards to noise, when noise is not present, the received packets are assumed to be delivered with no error with probability 1, while when noise is present a fixed bit error rate (BER) of 0.0001 is considered. Scenario 1, for example, corresponds to a network having 5 nodes with a low traffic load that represents an under-saturated network while noise is absent. Table~\ref{blocks} includes the blocks and their associated algorithm, mechanism, or parameters that are used by DeepMAC framework for the experiments. Some blocks have different algorithms or parameters. Table~\ref{blocks} also shows the default values for these blocks. As an example, if the fragmentation block is not selected by the agent, then the frame size remains 1500 bytes in the corresponding scenario. Otherwise if selected, the frame size varies. In order to see what blocks are selected by the agent in different scenarios,  the evaluation for each scenario is performed 20 times. We then collected those blocks that are selected together more frequently than others by the agent over 20 rounds of repeating each scenario.

\begin{table}[ht]

\footnotesize
\caption{Selected Blocks by DeepMAC Under Different Networking Load  \label{table:selected_blocks}}
\vspace{-3 mm}
\begin{tabular}{|c|c|c|}
\hline
\textbf{\# of Nodes} & \textbf{\begin{tabular}[c]{@{}c@{}}Traffic\\ Load\end{tabular}} & \textbf{Blocks Selected by DeepMAC}                                                                                                                                        \\ \hline
3                             & Low                                                             & No ACK, Aggregation                                                                                                                                           \\ \hline
15                              & Average                                                          & ACK, Fragmentation, BEB, CW                                                                                                                                                   \\ \hline
45                              & High                                                            & \begin{tabular}[c]{@{}c@{}}EIED, CW, CS, ACK, Fragmentation, RTS/CTS \end{tabular} \\ \hline
\end{tabular}
\tiny
\centering
\end{table}

\subsection{DeepMAC Against Baseline CSMA/CA}

We compare the performance of DeepMAC against CSMA/CA with a different number of contending nodes. CSMA uses ACK mechanism for successful transmissions. It also uses a Binary Exponential Backoff (BEB) technique to randomize each node attempt of transmitting, to reduce collision probability. CSMA/CA random backoff is decentralized and unable to efficiently handle collisions. Therefore, the network throughput degrades when the number of competing nodes increases. In the case of wireless CSMA/CA since collisions cannot be detected two main mechanisms to determine the successful reception of a frame are exchanging ACK and RTS/CTS control packets. The RTS/CTS mechanism can effectively ameliorate the hidden node problem. Although exchanging these control packets is useful for successful packet transmission, they could introduce extra overhead for bandwidth utilization.  In this experiment, we increase the number of contending nodes from 1 to 50 with the assumption of no noise. Such an assumption helps to simplify the effect of \textit{avalanche rate} when multiple transmission rates are available. Similarly, the traffic load increases from low to high.

\begin{table*}[t]

\begin{minipage}[t]{0.3\linewidth}\centering
\footnotesize
\centering
\caption{Simulation Configuration}
\begin{tabular}{c|c}
\textbf{Parameters}        & \textbf{Values}              \\ \hline
Frame Size                 & 1500Bytes (Default) \\
Time Slot                  & 0.2 msec                     \\
Channel Capacity           & 10 Mbps                      \\
Learning Rate ($\alpha$)   & 1                            \\
History Length ($H_t$)     & 15                           \\
Discount Factor ($\gamma$) & 0.8                         
\end{tabular}
\label{table_sim}
\end{minipage}\hfill %
\begin{minipage}[t]{0.7\linewidth}\centering
\centering
\caption{Blocks and their associated algorithm/ mechanism/ parameter \label{blocks}}
\begin{tabular}{|c|c|c|}
\hline
\textbf{Building Block} & \textbf{Algorithm /  Parameter}                                                        & \textbf{Default}           \\ \hline
Backoff                 & BEB, EIED                                                                              & BEB                        \\
ACK                     & \begin{tabular}[c]{@{}c@{}}No ACK,  ACK\end{tabular} & ACK                        \\
Fragmentation (Fr)           & Packet Size =200, 500, 1000 bytes                                                           & Packet Size = 1500 bytes   \\
Aggregation (Ag)             & Packet Size =2000 bytes                                                                & Packet Size =1500 bytes    \\
RTS/CTS                 & Enabled/Disabled                                                                       & N/A                    \\
CW       & 0-1023                                                                                 & $CW_{min} = 15$ \\
Carrier Sense (CS)      & Enabled/Disabled                                                                       & N/A                    \\
Data Transmission Rate (DR)  & 6/9/12/24/36/48/54 (Mbps)                                                              & 54 Mbps                    \\ \hline
\end{tabular}
\end{minipage}
\end{table*}

\begin{table*}[]
\begin{minipage}[t]{0.3\linewidth}\centering
\footnotesize
\caption{Simulation scenarios}
\label{table_scenario}
\begin{tabular}{|c|c|c|c|}
\hline
Scenario  & Nodes & Load & Noise \\ 
\hline
1                 & 5    & Low  & No    \\
\hline
2                & 5   & Low  & Yes   \\
\hline
3              & 15  & Average  & No    \\
\hline
4              & 15   & Average  & Yes   \\
\hline
5                 & 20  & High & No    \\
\hline
6               & 20  & High & Yes   \\
\hline
7              & 50  & Saturated & No    \\
\hline
8              & 50  & Saturated & Yes   \\ 
\hline
\end{tabular}
\end{minipage}\hfill %
\begin{minipage}[t]{0.7\linewidth}\centering
\footnotesize
\caption{Blocks selected by DeepMAC agent}
\label{table_components}
\begin{tabular}{|c|
>{\columncolor[HTML]{FFCCC9}}c |c|c|c|
>{\columncolor[HTML]{FFCCC9}}c |c|
>{\columncolor[HTML]{FFCCC9}}c |c|c|c|}
\hline
\#   & DR                        & BEB                                             & EIED                     & CS                                              & \cellcolor[HTML]{FFFFFF}CW & No ACK                   & \cellcolor[HTML]{FFFFFF}ACK & Fr                           & Ag                           & RTS/CTS                  \\ \hline
1 & {\color[HTML]{000000} 54} &                                                 &                          &                                                 & 31                         & \cellcolor[HTML]{FFCCC9} & \cellcolor[HTML]{FFFFFF}    &                              & \cellcolor[HTML]{FFCCC9}2000 &                          \\ \hline
2 & 24                        &                                                 &                          & \cellcolor[HTML]{FFCCC9}                        & 31                         & \cellcolor[HTML]{FFCCC9} & \cellcolor[HTML]{FFFFFF}    &                              &                              &                          \\ \hline
3 & 54                        &                                                 &                          & \cellcolor[HTML]{FFCCC9}                        & \cellcolor[HTML]{FFFFFF}   &                          &                             &                              &                              &                          \\ \hline
4 & 48                        &                                                 & \cellcolor[HTML]{FFCCC9} & \cellcolor[HTML]{FFFFFF}{\color[HTML]{FFFFFF} } & 15                         &                          &                             &                              &                              & \cellcolor[HTML]{FFCCC9} \\ \hline
5 & 54                        &                                                 & \cellcolor[HTML]{FFCCC9} & \cellcolor[HTML]{FFCCC9}                        & 15                         &                          &                             &                              &                              & \cellcolor[HTML]{FFCCC9} \\ \hline
6 & 24                        & \cellcolor[HTML]{FFCCC9}{\color[HTML]{333333} } &                          &                                                 & 15                         &                          &                             & \cellcolor[HTML]{FFCCC9}1000 &                              &                          \\ \hline
7 & 36                        & \cellcolor[HTML]{FFCCC9}                        &                          &                                                 & 15                         &                          &                             & \cellcolor[HTML]{FFCCC9}500  &                              & \cellcolor[HTML]{FFCCC9} \\ \hline
8 & 24                        & \cellcolor[HTML]{FFCCC9}                        &                          &                                                 & 15                         &                          &                             & \cellcolor[HTML]{FFCCC9}500  &                              & \cellcolor[HTML]{FFCCC9} \\ \hline

\end{tabular}
\end{minipage}
\end{table*}

In the following, we describe the throughput gains of DeepMAC against CSMA/CA for two Different traffic loads: Low and High \footnote {It would be interesting to compare the performance for other objectives as fairness and latency. However, since our objective function only optimizes average throughput, we consider only the throughput for this experiment.}. In the following, we try to get reasonable insights about what did DeepMAC additionally learned to dominate CSMA/CA.

\textbf{Low traffic load}
 In the first experiment, illustrated in Figure~\ref{fig:lowload}, the number of contending nodes increases from 1 to 15. In our simulation, every 3 seconds a new node joins the network, and the simulation duration lasts for 45 seconds. As illustrated by Figure~\ref{fig:lowload}, CSMA/CA fails to fully utilize the channel bandwidth, while DeepMAC protocol effectively adapts to the network load changes by selecting the appropriate set of building blocks. Intuitively, DRL agent  learns that when the load of the network is low, it could eliminate control packets e.g., ACK, to increase the throughput of the channel. In Table~\ref{table:selected_blocks}, we closely looked at the blocks selected by DeepMAC when the number of nodes is 3. There is an interesting observation of the selected blocks. The No ACK mechanism is selected along with Aggregation that both can enhance the throughput by reducing extra control frame overhead. Intuitively, DeepMAC has learned what blocks are important to choose for this scenario. These observations may look intuitive for a human, but it makes it interesting when a DRL agent is able to learn such intuition on its own.

\textbf{High traffic load} 
In the second experiment, we consider a high load traffic network where at the start of the experiment 25 nodes are competing for the channel. We then add nodes every two seconds until the number of contending nodes reaches 50 as shown in Figure~\ref{fig:highload}. By looking at Table~\ref{table:selected_blocks}, we observe that DeepMAC has selected EIED (Exponential Increase Exponential Decrease) over BEB. This can be because the slower reduction rate helps improving saturation throughput. Besides control packets are also selected by the agent probably to avoid the collisions and retransmission of large data packets.

\subsection{Selected Blocks in Different Scenarios}
\label{important-blocks}

 This subsection focuses on the selected blocks by the agent in different scenarios. The selected blocks are shown in Table~\ref{table_components} in which the pink color cells indicate the active blocks in each scenario along with their selected values (If the block has a tunable parameter), while white cells indicate that the corresponding block is inactive in the given scenario. In the following, we divide our observations about DeepMAC behavior in three parts and discuss further  about each case individually.

\textbf{Low load with/without noise}
In scenarios with the low load when the noise is absent (e.g., Scenario \#1). As we can see in Table~\ref{table_components} cells corresponding to control packets such as ACK or RTS/CTS are inactive by the agent.  This observation is justifiable. Even though the control packets are much smaller than the data packets, the time spent for control packet transmission is not negligible.  

Therefore, when the network is under saturated, and the number of competing nodes are small, the DRL agent avoids control packet overheads to maximize the throughput. Intuitively, to reduce the relative percentage of the time loss due to packet overhead and MAC coordination, frame aggregation is also selected by the agent. While for the same scenario, when the noise is present, it adds Career Sensing (CS) block. This may be due to the fact that the agent learns such a mechanism could be useful when the throughput drops.  

\textbf{Average load with/ without noise} For scenarios with the average level of noise (Scenario \#3 and \#4) except common ACK mechanism selection, there is no obvious pattern. This observation could be either because such scenarios are not able to capture the useful information of what specific blocks should be selected, or it is simply because selecting different blocks does not provide a significant difference in the achieved throughput in such scenarios. 

\textbf{High and saturated load with/without noise} We divide  our observations for the three following scenarios: (1) The first observation in the high and saturated scenarios (Scenario \#5 to \#8) is the ACK mechanism is selected by the agent. Intuitively, this could be because the agent learns such a mechanism can contribute to prevent more number of collisions and retransmissions to enhance the throughput. (2) When comparing scenario 5 to 6, we observe that the agent activates the Fragmentation block. The size of the sub-frames in practice plays an important factor that can influence network throughput performance for a given channel condition. The larger packets could contribute to the higher Packet Error Rate (PER) which would cause throughput drop due to a large number of retransmissions. (3) When the network is saturated, the agent selects protection mechanisms such as ACK and RTS/CTS along with smaller frame sizes and lower bitrate. However, it is not clearly obvious if the smaller frames contribute much to enhance the throughput. This is due to the fact that small fragments with the extra introduced overhead could also decrease the throughput performance.

The varying results reveal why it is extremely hard for an algorithm based on manually-specified rules and thresholds to capture the optimal solution, and why instead it is helpful to use machine-learning techniques to optimize the design of control algorithms as well as,  getting insights about what functionality (block) is useful under what scenario.

\section{Discussion and Conclusion}
 
In this paper, we proposed and evaluated a framework for MAC protocol design optimization using a DRL-based approach. We have shown that by observing the decisions of the DeepMAC agent and using a method such as input modularization (protocol decomposition into building blocks), it is possible to extract information about the associated component selection by the agent. We envision this method could offer useful insights, especially to protocol designers to build a deeper perception about the significance of an individual or a set of protocol blocks (functions) under different scenarios. This could help them focusing on enhancements/ modifications of important protocol components than focusing on the whole protocol performance which can contribute to enhancing the overall protocol design and performance. However, more work and a deeper analysis need to be done to discuss additional issues specific to understanding DRL models.

In our proposed framework we considered a centralized agent that is considered as a supernode that learns the best combination set of blocks and enforces this set to the rest of the nodes in the network. However, one of the main challenges is to design a distributed multi-agent DRL algorithm in which each node works locally towards serving its own objective. This is a challenging open direction and requires DeepMAC approach to be extended to incorporate fairness criteria. Another open question is whether it would be beneficial to combine blocks from different medium access methods e.g., TDMA, CSMA/CA, etc., available to the agent to select from while designing protocols, especially in heterogeneous environments.

\renewcommand{\refname}{\footnotesize References}
\bibliographystyle{abbrv}
{\footnotesize
\bibliography{references}}

\begin{thebibliography}{10}

\bibitem{bai:2004}
F.~Bai, G.~Bhaskara, and A.~Helmy.
\newblock Building the blocks of protocol design and analysis: challenges and
  lessons learned from case studies on mobile ad hoc routing and micro-mobility
  protocols.
\newblock {\em ACM SIGCOMM Computer Communication Review}, 34(3):57--70, 2004.

\bibitem{pasandi2019poster}
H.~Barahouei~Pasandi and T.~Nadeem.
\newblock Poster: Towards self-managing and self-adaptive framework for
  automating mac protocol design in wireless networks.
\newblock In {\em Proceedings of the 20th International Workshop on Mobile
  Computing Systems and Applications}, pages 171--171. ACM, 2019.

\bibitem{dsa3}
U.~Challita, L.~Dong, and W.~Saad.
\newblock Proactive resource management in lte-u systems: A deep learning
  perspective.
\newblock {\em arXiv preprint arXiv:1702.07031}, 2017.

\bibitem{ieee:2016}
I.~C. S. L. M.~S. Committee et~al.
\newblock Part 11: Wireless lan medium access control (mac) and physical layer
  (phy) specifications.
\newblock 2016.

\bibitem{doerr:2005}
C.~Doerr, M.~Neufeld, J.~Fifield, T.~Weingart, D.~C. Sicker, and D.~Grunwald.
\newblock Multimac-an adaptive mac framework for dynamic radio networking.
\newblock In {\em First IEEE International Symposium on New Frontiers in
  Dynamic Spectrum Access Networks, 2005. DySPAN 2005.}, pages 548--555. IEEE,
  2005.

\bibitem{eini:2019}
R.~Eini and S.~Abdelwahed.
\newblock Distributed model predictive control for intelligent traffic system.
\newblock In {\em 2019 International Conference on Internet of Things (iThings)
  and IEEE Green Computing and Communications (GreenCom) and IEEE Cyber,
  Physical and Social Computing (CPSCom) and IEEE Smart Data (SmartData)},
  pages 909--915. IEEE, 2019.

\bibitem{geyer:2018}
F.~Geyer and G.~Carle.
\newblock Learning and generating distributed routing protocols using
  graph-based deep learning.
\newblock In {\em Proceedings of the 2018 Workshop on Big Data Analytics and
  Machine Learning for Data Communication Networks}, pages 40--45. ACM, 2018.

\bibitem{sanitize:2017}
J.~Huang, Q.~Li, S.~Zhong, L.~Liu, P.~Zhong, J.~Wang, and J.~Ye.
\newblock Synthesizing existing csma and tdma based mac protocols for vanets.
\newblock {\em Sensors}, 17(2):338, 2017.

\bibitem{deepcc}
N.~Jay, N.~H. Rotman, P.~Godfrey, M.~Schapira, and A.~Tamar.
\newblock Internet congestion control via deep reinforcement learning.
\newblock {\em arXiv preprint arXiv:1810.03259}, 2018.

\bibitem{piva:2019}
G.~Maselli, M.~Piva, and J.~A. Stankovic.
\newblock Adaptive communication for battery-free devices in smart homes.
\newblock {\em IEEE Internet of Things Journal}, 6(4):6977--6988, 2019.

\bibitem{dsa}
O.~Naparstek and K.~Cohen.
\newblock Deep multi-user reinforcement learning for distributed dynamic
  spectrum access.
\newblock {\em IEEE Transactions on Wireless Communications}, 18(1):310--323,
  2018.

\bibitem{panahi:2019}
A.~Panahi, S.~Saeedi, and T.~Arodz.
\newblock word2ket: Space-efficient word embeddings inspired by quantum
  entanglement.
\newblock {\em arXiv preprint arXiv:1911.04975}, 2019.

\bibitem{pasandi-b:2019}
H.~B. Pasandi.
\newblock Towards a machine learning-based framework for automated design of
  networking protocols.
\newblock In {\em 2019 IEEE International Conference on Pervasive Computing and
  Communications Workshops (PerCom Workshops)}, page 433–434. IEEE, 2019.

\bibitem{pasandi:2019}
H.~B. Pasandi and T.~Nadeem.
\newblock Challenges and limitations in automating the design of mac protocols
  using machine-learning.
\newblock In {\em 2019 International Conference on Artificial Intelligence in
  Information and Communication (ICAIIC)}, pages 107--112. IEEE, 2019.

\bibitem{mac-unboxing:2020}
H.~B. Pasandi and T.~Nadeem.
\newblock Unboxing mac protocol design optimization using deep learning.
\newblock In {\em 2020 IEEE International Conference on Pervasive Computing and
  Communications Workshops (PerCom Workshops)}. IEEE, 2020.

\bibitem{dmdl}
P.~Wang, M.~Petrova, and P.~M{\"a}h{\"o}nen.
\newblock Dmdl: A hierarchical approach to design, visualize, and implement mac
  protocols.
\newblock In {\em Wireless Communications and Networking Conference (WCNC),
  2018 IEEE}, pages 1--6. IEEE, 2018.

\bibitem{dsa2}
S.~Wang, H.~Liu, P.~H. Gomes, and B.~Krishnamachari.
\newblock Deep reinforcement learning for dynamic multichannel access in
  wireless networks.
\newblock {\em IEEE Transactions on Cognitive Communications and Networking},
  4(2):257--265, 2018.

\bibitem{wic:2017}
C.~S. Wickramasinghe, K.~Amarasinghe, and M.~Manic.
\newblock Parallalizable deep self-organizing maps for image classification.
\newblock In {\em 2017 IEEE Symposium Series on Computational Intelligence
  (SSCI)}, pages 1--7. IEEE, 2017.

\bibitem{wick:2018}
C.~S. Wickramasinghe, D.~L. Marino, K.~Amarasinghe, and M.~Manic.
\newblock Generalization of deep learning for cyber-physical system security: A
  survey.
\newblock In {\em IECON 2018-44th Annual Conference of the IEEE Industrial
  Electronics Society}, pages 745--751. IEEE, 2018.

\bibitem{wic:2019}
C.~S. Wicramasinghe, K.~Amarasinghe, and M.~Manic.
\newblock Deep self-organizing maps for unsupervised image classification.
\newblock {\em IEEE Transactions on Industrial Informatics}, 2019.

\bibitem{dlma}
Y.~Yu, T.~Wang, and S.~C. Liew.
\newblock Deep-reinforcement learning multiple access for heterogeneous
  wireless networks.
\newblock {\em IEEE Journal on Selected Areas in Communications},
  37(6):1277--1290, 2019.

\bibitem{mac_model:2004}
H.~Zhai, Y.~Kwon, and Y.~Fang.
\newblock Performance analysis of ieee 802.11 mac protocols in wireless lans.
\newblock {\em Wireless communications and mobile computing}, 4(8):917--931,
  2004.

\bibitem{queue:2012}
S.~Zhuo, Y.-Q. Song, Z.~Wang, and Z.~Wang.
\newblock Queue-mac: A queue-length aware hybrid csma/tdma mac protocol for
  providing dynamic adaptation to traffic and duty-cycle variation in wireless
  sensor networks.
\newblock In {\em 2012 9th IEEE International Workshop on Factory Communication
  Systems}, pages 105--114. IEEE, 2012.

\end{thebibliography}
\end{document}